\documentclass[sigconf]{acmart}

\usepackage[inline]{enumitem}
\usepackage{multirow}

\copyrightyear{2026}
\acmYear{2026}
\setcopyright{cc}
\setcctype{by}
\acmConference[IH\&MMSec '26]{ACM Workshop on Information Hiding and Multimedia Security}{June 17--19, 2026}{Firenze, Italy}
\acmBooktitle{ACM Workshop on Information Hiding and Multimedia Security (IH\&MMSec '26), June 17--19, 2026, Firenze, Italy}
\acmDOI{10.1145/3785353.3815087}
\acmISBN{979-8-4007-2376-6/2026/06}

\newcommand{\vect}[1]{\mathbf{#1}}

\begin{document}

\title{Explainable-by-Design Audio Deepfake Detection via Wiener-Hopf Linear Prediction}

\author{Mattia Tamiazzo}
\orcid{0009-0009-1218-5659}
\email{mattia.tamiazzo.1@phd.unipd.it}
\affiliation{%
  \department{Department of Information Engineering}
  \institution{University of Padova}
  \city{Padova}
  \country{Italy}
}

\author{Simone Milani}
\orcid{0000-0001-8266-5839}
\email{simone.milani@dei.unipd.it}
\affiliation{%
  \department{Department of Information Engineering}
  \institution{University of Padova}
  \city{Padova}
  \country{Italy}
}

\author{Massimo Iuliani}
\orcid{0000-0002-5501-4667}
\email{massimo.iuliani@ampedsoftware.com}
\affiliation{
  \institution{Amped Software}
  \city{Trieste}
  \country{Italy}
}

\author{Marco Fontani}
\orcid{0000-0002-0630-2128}
\email{marco.fontani@ampedsoftware.com}
\affiliation{
  \institution{Amped Software}
  \city{Trieste}
  \country{Italy}
}

\settopmatter{authorsperrow=2}

\begin{abstract}
    The rapid advancement of synthetic speech generation methods has made audio deepfake detection a critical challenge in multimedia forensics.
    While recent approaches achieve high detection accuracy, they typically rely on black-box architectures that offer limited interpretability and high computational complexity.
    In this paper, we propose an explainable-by-design audio deepfake detection framework based on Wiener-Hopf linear prediction, processed by a lightweight 2D Convolutional Neural Network (CNN). 
    This design enables a direct and transparent connection between classification outcomes and the acoustic properties of the signal.
    Experimental results on benchmark datasets demonstrate competitive detection performance while maintaining significantly lower computational complexity compared to state-of-the-art solutions.
    The interpretability analysis using Grad-CAM reveals that the classifier focuses on low-order predictor coefficients and on silence and transitional regions, suggesting that the Wiener-Hopf predictor captures reverberation characteristics and subtle statistical inconsistencies in synthetic speech.
    Finally, robustness experiments show that fine-tuning effectively recovers detection performance under common post-processing degradations, including additive noise, MP3 compression, and telephone filtering.
\end{abstract}



\maketitle

\section{Introduction}
In recent multimedia forensics research, a significant effort has been dedicated to the design of effective synthetic speech detectors \cite{10.1145/3643491.3660289,Yi2023AudioDD}, generating a wide range of processing schemes and model architectures, ranging from statistical and sound properties~\cite{mari2022sound} to attention schemes~\cite{wifs2025attention}. 
Even though such solutions have shown increasing accuracy over the recent years, there are still many issues that remain unresolved~\cite{openADD}.

One of the main issues is adaptability: the performance of existing methods can be very good on controlled and stable scenarios, but rapidly degrades as datasets and acquisition conditions change. 
To mitigate this, continual learning~\cite{salvi2025freeze} and domain adaptation strategies~\cite{10286049} have been proposed, together with Mixture-of-Experts~\cite{wifs2025attention} and feature fusion~\cite{Yang2024ARA}. 
Indeed, this problem is mainly due to the adoption of large classification networks which tend to adapt and overfit to specific dataset conditions. 
Moreover, model sizes make their adoption in embedded systems more difficult due to complexity constraints.

A second issue concerns the explainability of detector outcomes.
Most of the proposed solutions resort to learned features that permit achieving top accuracy metrics but fail to provide adequate justifications. 
For such reasons, recent works have investigated eXplainable AI (XAI) solutions for audio deepfake detection~\cite{MOMIN2025105738,10887568}, which aim at highlighting those parts of the audio sequence that mostly contribute to the final decision.
Despite this, the involved detectors remain very complex, thus making the justification and description of the whole decision chain very hard.

The approach presented in this paper targets both justifiability and complexity following an \emph{explainability-by-design} architecture~\cite{10.1145/3708504}, where decisions can be directly related to hand-crafted and interpretable features by back-propagating outcomes through a low-complexity network. 

Based on the hypothesis that real and synthetic speech may exhibit different predictability characteristics (due to windowing or inconsistencies in propagations), the input audio sequence is predicted using a Wiener-Hopf estimation strategy: then, filter taps are stacked into a 2D matrix processed by a small Convolutional Neural Network (CNN). 
Final decisions are supported by Grad-CAM~\cite{selvaraju_grad-cam_2020}
heatmaps revealing the most distinctive traces and evaluating the reliability of the outcome.
Experimental results show that the proposed solution achieves competitive performances with respect to state-of-the-art works on heterogeneous datasets, while requiring a significantly lower computational effort.

The main contributions and novelties of the work can be summarized as follows.

\begin{itemize}
    \item The paper presents a new lightweight \emph{explainable-by-design} scheme for synthetic audio detection, where  Wiener-Hopf filter coefficients are  processed by means of a Convolutional Neural Network (CNN) and connected to a saliency map generated by Grad-CAM. Indeed, this is one of the first approaches that combines physically-relevant audio features with explainability algorithms at a limited computational cost while achieving the same performance as more complex schemes.
    \item The detector's outcomes are justified by a Grad-CAM heatmap revealing hard-to-predict audio segments that are characteristic of real speech.
    \item The approach provides competitive performance across heterogeneous datasets (ASVspoof 2019, Fake-or-Real, DiffSSD), thus showing its ability to adapt to different settings.
    \item Experimental tests were re-run under different real-world adversarial conditions to test the robustness of the approach; results show that, under proper training, no significant performance degradation is observed.
    \item The required computational complexity is $20$ times smaller than other state-of-the-art works with similar detection performance. 
\end{itemize}

The paper is organized as follows. 
Section~\ref{sec:related} reviews some of the recent literature works on synthetic speech detection, focusing on interpretability and explainability of features. Section~\ref{sec:method} describes the proposed method, while Section~\ref{sec:setup} presents the adopted experimental setting. 
Section~\ref{sec:results} reports the detection accuracy and highlights the revealing elements in the detection, including the interpretability study based on Grad-CAM and the robustness evaluation under post-processing operations. Finally, Section~\ref{sec:conclusions} summarizes the main findings and outlines future research directions.

\section{Related works}\label{sec:related}
    Synthetic speech detection has been widely investigated in recent years, with significant progress in feature extraction strategies, classification architectures, and benchmark datasets~\cite{li2025survey,Yi2023AudioDD}. 
    Many recent approaches rely on end-to-end deep learning pipelines that process raw waveforms or spectrograms. 
    More recently, large Self-Supervised Learning (SSL) models, such as Wav2Vec2~\cite{baevski_wav2vec_2020}, have been adopted as deep feature extractors, replacing traditional hand-crafted acoustic descriptors~\cite{li2025survey, tak_automatic_2022,9456037}. 
    Multimodal approaches are also worth mentioning whenever audio traces are supported by visual information; such solutions rely on discrepancies between video and sound or unnatural behaviors \cite{10354308,resemble2024multimodal}.
    Although they often display a high accuracy, their performance seems to be related to an aggregation of revealing traces rather than to explanatory characteristics in real speech.
    Finally, recent research efforts have also focused on ensuring a high accuracy under different acquisition conditions \cite{deepsonar} or datasets \cite{salvi2025freeze,negroni2025leveraging}. For this reason, the latest papers are concerned with continual learning and domain adaptation techniques that improve generalization across different datasets and acquisition conditions.
    
    These approaches achieve strong detection performance, but they often rely on black-box representations whose internal structure is difficult to interpret in forensic scenarios, and their large computational complexity may limit the deployment in resource-constrained settings. 

    To address these limitations, a parallel line of research has explored physically grounded acoustic representations, to improve generalization and interpretability.   
    Several works employ formant-based features to reveal artifacts introduced by neural vocoders and generative models. 
    Cuccovillo et al.~\cite{cuccovillo_audio_2023} proposed a spectrogram transformer based on formant magnitude and phase information, showing that phonetic features enhance both detection performance and interpretability.
    The work in~\cite{salvi_phoneme-level_2025} extended phoneme-level analysis to person-of-interest scenarios, showing that articulatory cues provide discriminative information for speaker-specific deepfake detection.
    Salvi et al.~\cite{salvi_towards_2023} analyzed the spectral regions that most influence detector decisions by applying XAI techniques including LIME~\cite{ribeiro_why_2016} and Grad-CAM~\cite{selvaraju_grad-cam_2020}, showing that specific frequency bands carry highly discriminative cues.
    Mari et al.~\cite{mari2022sound} showed that first-digit statistical properties of silence regions provide effective low-complexity cues for synthetic audio detection, suggesting that non-speech passages carry forensically relevant information.
    Following this direction, this work adopts Wiener-Hopf filter coefficients as interpretable feature within an explainability-by-design architecture~\cite{10.1145/3708504}.
    This enables decisions to be traced back to the physical properties of the signal, while maintaining a low computational complexity.

\section{Method}\label{sec:method}
An explainable approach to audio classification implies adopting a feature representation that is deeply connected to signal peculiarities for synthetic speech. Previous works have shown silence regions are very difficult to synthesize \cite{mari2022sound}, while stationarity and reverberation are closely related to the natural propagation of sounds within recording environments \cite{DBLP:journals/ejisec/BorrelliBAST21}. Detecting physical discontinuities in such features (which can hardly be compensated by learned approaches processing single segments of signal) proves to be both effective and interpretable for human analysis. To this purpose, we resort to representing the analyzed audio tracks by means of Wiener-Hopf filter coefficients whose values exhibit oscillatory patterns when reverberant components cannot be properly estimated or when signal properties change abruptly due to windowing effects introduced by synthesis algorithms.

    A Wiener-Hopf linear predictor~\cite{levinson_wiener_1946,wiener_extrapolation_1949} of order M is applied to consecutive overlapping windows of 1024 samples, with a stride of 512 samples (50\% overlap). 
    The predictor coefficients $\vect{h} \in \mathbb{R}^M$ are obtained by solving the Wiener-Hopf equation: 
    \begin{equation}
        \vect{R} \cdot \vect{h} = \vect{p},
    \end{equation}
    where $\vect{R} \in \mathbb{R}^{M \times M}$ is the Toeplitz autocorrelation matrix of the input signal $x$, and $\vect{p} = \bigl[r(1),r(2),\dots,r(M)\bigl]^T$ is the vector of autocorrelation values, with $r(k) = \mathbb{E}\bigl[x[n] \cdot x[n-k]\bigl]$ the autocorrelation of the signal $x$ at lag $k$.   
    The coefficients are computed using the Cholesky decomposition, exploiting the positive-definite structure of $\vect{R}$, and finding the least-squares solution in degenerate cases. 
    The predictor estimates each sample $\hat{x}[n]$ as a linear combination of the $M$ preceding samples, as follows:
    \begin{equation}
        \hat{x}[n] = \sum_{k=1}^M \vect{h}[k] \cdot x[n-k].
    \end{equation}

\begin{figure}
        \centering
        \includegraphics[width=\linewidth]{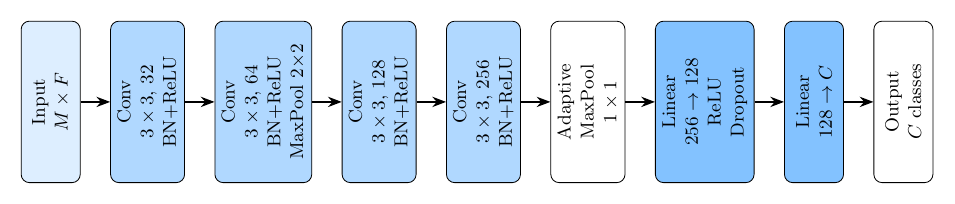}
        \caption{Architecture of the 2D CNN used in the work.}
        \label{fig:architecture}
        \Description{Block diagram of the 2D CNN used in the work, with four convolutional layers.}
    \end{figure}

    The coefficients $\vect{h}[k]$ of the Wiener-Hopf predictor extracted from each window are stacked across frames to form a 2D matrix of shape $(F \times M)$, where $F$ is the number of frames and $M$ is the predictor order. 
    As utterance length varies across audio samples, the matrices are padded to a fixed number of frames using symmetric padding, which avoids discontinuities at the boundaries. 
    The resulting matrix, treated as a single-channel 32-bit floating-point (float32) image, is passed to a simple 2D CNN classifier. 
    The network consists of four convolutional layers with batch normalization and ReLU activations, grouped in two blocks, with max pooling after the first block. 
    A global adaptive max pooling layer collapses the spatial dimensions, and the representation is passed to a fully connected classifier with dropout, which outputs the prediction logits for the classes present in the dataset. 
    The network architecture is shown in Figure~\ref{fig:architecture} and has a total of 422K trainable parameters. 

\section{Experimental setup}\label{sec:setup}

    This section describes the experimental protocol adopted to evaluate the proposed approach.
    We first introduce the considered datasets, followed by the training configuration and hyperparameter selection strategy.
    
    \subsection{Datasets}
        The experimental setup of our detection approach was defined considering several state-of-the-art audio deepfake datasets. 
        
        \begin{itemize}[leftmargin=0pt, label={}]
            \item \textbf{ASVspoof 2019 (ASV19)}~\cite{todisco_asvspoof_2019}.
            Introduced to assess countermeasures against synthetic and voice-converted  English speech, the dataset covers a wide range of generation systems, and is one of the most widely adopted benchmarks in the audio deepfake detection community. For our purposes, we focused on the Logical Access (LA) scenario. 
            \item \textbf{FakeOrReal (FoR)}~\cite{reimao_for_2019}.
            The dataset contains English bona fide speech, sourced from publicly available corpora, and synthetic speech generated by a variety of open-source and commercial systems. 
            \item \textbf{Diffusion-Based Synthetic Speech Dataset (DiffSSD)}~\cite{bhagtani_diffssd_2025}.
            This corpus comprises approximately 200 hours of labeled English speech, with synthetic samples generated by 8 open-source and 2 commercial diffusion-based Text-To-Speech (TTS) systems. The real speech data was sourced from LJSpeech~\cite{ito_lj_2017} and LibriSpeech~\cite{panayotov_librispeech_2015}. 
        \end{itemize}

    \subsection{Hyperparameter tuning}
        The Wiener-Hopf predictor order was set to $M = 30$, offering a good trade-off between prediction accuracy and computational complexity. 
        For the CNN classifier, the maximum number of frames was set to $F = 300$ using symmetric padding.
        This value covers the majority of utterances in the datasets without excessive padding. 
        To evaluate the impact of this design choice on detection performance, we also tested with standard zero-padding of the input.
        The model was trained using cross-entropy loss with the AdamW optimizer with a learning rate of $3 \times 10^{-4}$ and weight decay of $10^{-4}$.
        The learning rate was reduced by a factor of $0.5$ when the validation loss did not improve for $3$ consecutive epochs. 
        The training was performed for a maximum of $50$ epochs with early stopping using a patience of $10$, a batch size of $32$ and a dropout rate of $0.5$. 
        Hyperparameters were determined empirically by evaluating multiple configurations on the development set.

\section{Results and discussion}\label{sec:results} 

    This section presents the classification performance of the model and the analysis of the most discriminative elements in the detection process.
    Moreover, we evaluate the robustness of the model to data perturbations.
    The evaluation is performed on all considered datasets using both the Equal Error Rate (EER) and the Area Under the Curve (AUC) for the Receiver Operating Characteristic (ROC) curve.     
    
    \subsection{Deepfake detection performance}
        We tested the proposed model on the three evaluation benchmarks, using the same training, validation, and test splits defined in the original protocols.
        The detection performance is presented in Table~\ref{tab:performance_results}.  

        \begin{table*}[t]
            \centering
            \small
            \caption{EER (\%) and AUC (\%) values of models evaluated on different datasets. The performance of the classifier zero-padding the feature matrices is compared with the symmetric padding. $\uparrow$ means the higher the better. $\downarrow$ means the lower the better. Average results marked with (*) are computed excluding DiffSSD.}
            \label{tab:performance_results}
            \begin{tabular}{lccccccccc}
            \toprule
             & \multicolumn{2}{c}{\textbf{ASV19}} & \multicolumn{2}{c}{\textbf{DiffSSD}} & \multicolumn{2}{c}{\textbf{FoR}} & \multicolumn{2}{c}{\textbf{Average}} \\
            \cmidrule(lr){2-3} \cmidrule(lr){4-5} \cmidrule(lr){6-7} \cmidrule(lr){8-9}
            \textbf{Architecture} & EER $\downarrow$ & AUC $\uparrow$ & EER $\downarrow$ & AUC $\uparrow$ & EER $\downarrow$ & AUC $\uparrow$ & EER $\downarrow$ & AUC $\uparrow$ \\
            \midrule
            2DCNN Zero-padding & 11.65 & 94.69 & 3.83 & 99.26 & 8.68 & 97.14 & 8.05 & 97.03 \\
            2DCNN Symmetric padding & {5.97}  & {96.50} & {2.95} & {\bf 99.62} & {\bf 0.56} & {\bf 99.99} & {\bf 3.16} & {\bf 98.70} \\
            \midrule
            MoE \cite{negroni2025leveraging} & 9.45  & 96.17 & {\bf 2.53} & 94.52 & 2.74 & 99.43 & 4.91 & 96.71 \\
            Pyramid feat. with DS \cite{11226156} & 15.60 & 92.70 & 30.60 & 75.70 & 9.10 & 97.4 & 18.43 & 88.60 \\
            Rawnet2 \cite{9414234} & 10.60 & 91.90 & 44.90 & 56.30 & 14.80 & 93.70 & 23.43 & 80.63 \\
            Wav2Vec2 \cite{tak2022wav2vec2spoof} & {\bf 2.56} & {\bf 99.60} & 3.00 & 99.58 & 7.46 & 96.90 & 4.34 & 98.69 \\
            Rawformer \cite{xiao2025rawtfnetlightweightcnnarchitecture} & 3.95 & 98.20 & -- & -- &  4.41 & 95.60 & 4.18* & 96.90* \\
            LCNN \cite{wu_light_2018} & 11.10 & 95.70 & -- & -- & 4.20 & 99.00 & 7.65* & 97.35* \\
            \bottomrule
            \end{tabular}%
        \end{table*}

        The 2D CNN employing symmetric padding achieves competitive results across all datasets, yielding an average EER of 3.16\% and an AUC of 0.987.
        Generally speaking, the mentioned approach achieves the highest average accuracy with respect to other solutions presented during the last years \cite{negroni2025leveraging,11226156,9414234}.

        On ASVspoof 2019, the proposed system obtains an EER of 5.97\%, which outperforms the official challenge baselines~\cite{todisco_asvspoof_2019} and remains competitive with state-of-the-art approaches, many of which rely on complex ensemble strategies.
        On DiffSSD, our model achieves 2.95\% EER on the test set, performing better than all the baselines reported in~\cite{bhagtani_diffssd_2025}, including large semi-supervised transformer-based models, such as Wav2Vec2 (3.00\%) and MFCC-ResNet (11.00\%).
        On the FoR dataset, the proposed model reaches an EER of 0.56\%, matching the performance of other high-complexity systems reported in~\cite{dowerah_speech_2026}.

        Across all three datasets, the symmetric padding variant consistently outperforms the zero-padding counterpart, confirming that preserving boundary continuity is critical for robust feature extraction. 
        Overall, the proposed explainable-by-design architecture provides a physically grounded and interpretable alternative to black-box detectors, while maintaining competitive detection performance across diverse evaluation benchmarks.

    \subsection{Interpretability analysis} \label{subsec:interpret}
\begin{figure*}[t]
            \centering
            \includegraphics[width=0.935\textwidth]{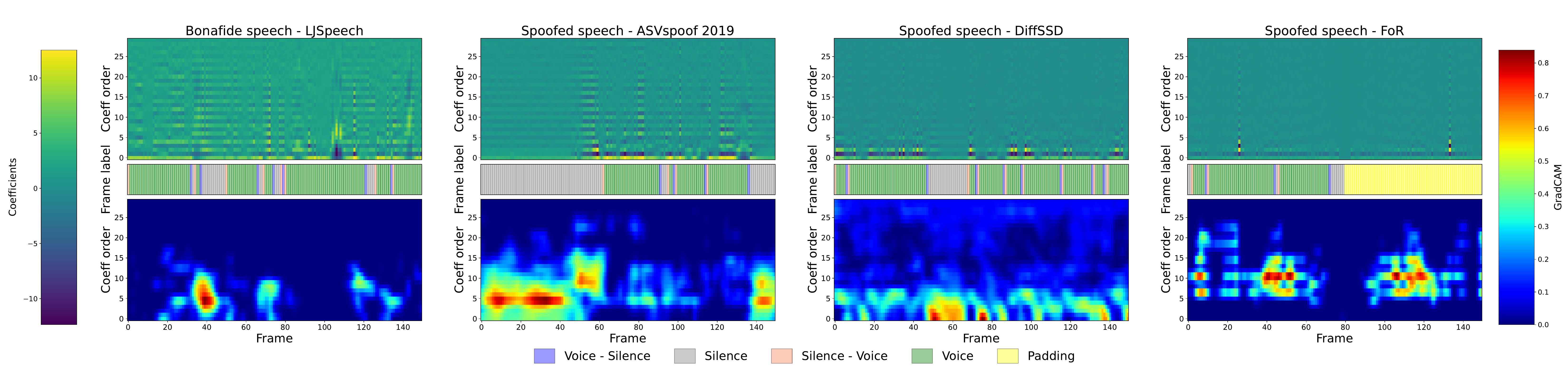}
            \caption{Grad-CAM analysis of representative samples from the datasets. Each column shows the Wiener-Hopf coefficient matrix (top image) for each frame, the corresponding voiced/unvoiced label (middle image), and the Grad-CAM saliency map (bottom image). High activations concentrate on low-order coefficients and are frequently observed in silence and transition regions.}
            \label{fig:gradcam}
            \Description{}
        \end{figure*}

        To better understand the discriminative behavior of the proposed approach, we apply Gradient-weighted Class Activation Mapping (Grad-CAM)~\cite{selvaraju_grad-cam_2020} to the last convolutional layer of the classifier. 
        Grad-CAM produces a saliency map over the input matrix, highlighting which Wiener-Hopf coefficients and temporal frames are the most informative for the classification decision. 
        Some examples of the generated images are presented in Figure~\ref{fig:gradcam}.
        
        The analysis reveals that the activations concentrate on low-order predictor coefficients, which primarily encode the coarse spectral envelope of the signal, characterizing the primary formant resonances of the vocal tract and approximating its gross geometry under the acoustic tube interpretation of linear prediction~\cite{makhoul_linear_1975,wakita_direct_1973}. 
        
        To characterize the temporal distribution of the activations, we classify each frame as voiced or unvoiced based on the short-time energy (STE) of the signal. 
        In particular, a frame is labeled as voiced if its energy exceeds $5\%$ of the maximum RMS energy of the utterance, and unvoiced otherwise.
        We observe that high activations of the network are frequently present in silence regions and on transitions between voiced and unvoiced passages.
        Our observations are consistent with previous studies~\cite{sivaraman2025investigating,negroni2026multi}, which report that models analyzing synthetic speech rely predominantly on unvoiced regions, suggesting that the most salient artifacts are concentrated in these segments. 
        In contrast, \cite{10887568} reports that Grad-CAM applied to a Wav2Vec2-AASIST model on ASVspoof 2019 highlights speech regions as the most discriminative.
        This discrepancy may be related to the different feature space: while the referenced model operates on raw waveforms, our approach focuses on Wiener-Hopf prediction coefficients, which are sensitive to the statistical structure of silence passages.
        We hypothesize that this behavior reflects the ability of the Wiener-Hopf predictor to capture features related to the reverberation effects of the recording environment. 
        In fact, while natural speech recordings typically exhibit reverberation tails in silent passages, deepfake generators tend to generate clean speech that lacks this acoustic pattern. 
        The linear predictor may therefore expose the absence of natural reverberation as a discriminative cue, providing a physically-based interpretation of the detection performance of the model.
    
    \subsection{Ablation study}
        To further investigate the contribution of different temporal regions to the detection performance, we conduct an ablation study by selectively removing voiced and silence frames (thus, generating \textit{no-voice} and \textit{no-silence} configurations, respectively).
        For each configuration, the Wiener-Hopf method is applied to the entire original audio excerpt to avoid discontinuities, discarding the coefficient vectors corresponding to the removed frames; the resulting matrices are then used to retrain the CNN classifier from scratch.
        
        The results of the ablation study are presented in Table~\ref{tab:ablation}.
        As expected, removing silence frames leads to a significant degradation in performance (increasing the EER by 14.42 \% on average), confirming the importance of non-speech segments in the detection process. 
        On the other hand, the exclusion of voiced frames increases the EER by 2.89 \% on average, preserving good detection performance. 
        These findings are consistent with the Grad-CAM analysis presented in Section \ref{subsec:interpret},
        and further support the importance of voice-to-silence and silence-to-voice passages in the detection (where Wiener-Hopf coefficients highlight inconsistencies at transition boundaries).

    \begin{table}[t]
        \centering
        \small
        \caption{EER (\%) considering modified versions of the datasets, where the model is trained and tested on the original data, or on versions with voiced or silence frames removed.}
        \label{tab:ablation}
        \setlength{\tabcolsep}{4pt}
        \begin{tabular}{lcccc}
            \toprule
            \textbf{Data condition} & \textbf{ASV19} & \textbf{DiffSSD} & \textbf{FoR} & \textbf{Average} \\
            \midrule
            Original data & 5.97 & 2.95 & 0.56 & 3.16 \\
            \textit{No-voice} & 6.35 & 6.83 & 4.96 & 6.05 \\
            \textit{No-silence} & 29.91 & 5.46 & 17.38 & 17.58 \\
            \bottomrule
        \end{tabular}
    \end{table}

    \subsection{Robustness to post-processing}
        To evaluate the robustness of the proposed method to real-world post-processing conditions, we apply a set of common signal degradations and measure the resulting detection performance.
        We considered additive noise, introduced using white, pink, and brown noise at four Signal-to-Noise Ratio (SNR) levels: 20, 15, 10, and 5 dB.
        Secondly, MP3 compression is applied at different bitrates (32, 64, 128 kbps), to simulate lossy compression artifacts commonly introduced during audio sharing and storage. 
        Moreover, bandpass filtering is applied using two frequency-range configurations (300-3400 Hz, 500-3000 Hz) in order to simulate the bandwidth limitations of telephone and VoIP transmissions.

        Table \ref{tab:robustness} reports the EER (\%) of the proposed model under the considered perturbations, evaluated in two conditions: frozen model (F), where the classifier trained on clean data is tested on the perturbed test set, and full network fine-tuning (N), where the model is retrained on the perturbed training data.
        The fine-tuning of the model is performed considering the original learning rate reduced by a factor of $0.1$, to limit the performance degradation on the original data. 
        As expected, the frozen model suffers a significant performance degradation under most perturbations, in particular under white noise and telephone filtering, with an EER exceeding 50 \% at low SNR levels. 
        In contrast, the addition of pink and brown noise causes a more moderate degradation.
        Fine-tuning consistently recovers most of the lost performance across all perturbation types and datasets, with EER values close to the clean baseline in most conditions. 
        Remarkably, on FakeOrReal the fine-tuned model achieves very low EER under MP3 compression and telephone filtering, suggesting that the discriminative cues captured by the Wiener-Hopf predictor are partially preserved even after a considerable bandwidth reduction.
        
        \begin{table}[t]
            \footnotesize
            \centering
            \caption{EER (\%) under various perturbations and retraining strategies across the datasets. F = Frozen model, N = Full network fine-tuning. Bold values indicate the best result between F and N for each condition.}
            \label{tab:robustness}
            \vspace*{-2ex}
            \setlength{\tabcolsep}{4pt}
            \begin{tabular}{ll cc cc cc}
                \toprule
                & & \multicolumn{2}{c}{\textbf{ASV19}} & \multicolumn{2}{c}{\textbf{DiffSSD}} & \multicolumn{2}{c}{\textbf{FoR}} \\
                \cmidrule(lr){3-4} \cmidrule(lr){5-6} \cmidrule(lr){7-8}
                \textbf{Perturbation} & \textbf{Condition} & F & N & F & N & F & N \\
                \midrule
                No perturbation & Clean data & \multicolumn{2}{c}{5.97} & \multicolumn{2}{c}{2.95} & \multicolumn{2}{c}{0.56} \\
                \midrule
                \multirow{3}{*}{MP3}
                    & 128 kbps & 32.90 & \textbf{11.39} & 22.68 & \textbf{7.10} & 13.25 & \textbf{0.17} \\
                    & 64 kbps  & 33.43 & \textbf{11.60} & 23.47 & \textbf{7.01} & 12.21 & \textbf{0.22} \\
                    & 32 kbps  & 34.59 & \textbf{11.92} & 18.15 & \textbf{7.32} & 27.86 & \textbf{0.35} \\
                \midrule
                \multirow{4}{*}{White Noise}
                    & SNR = 20 dB & 63.51 & \textbf{9.02}  & 45.44 & \textbf{3.93} & 52.89 & \textbf{13.55} \\
                    & SNR = 15 dB & 54.70 & \textbf{9.84}  & 49.19 & \textbf{4.18} & 71.26 & \textbf{14.22} \\
                    & SNR = 10 dB & 46.39 & \textbf{10.81} & 52.66 & \textbf{5.00} & 76.03 & \textbf{13.66} \\
                    & SNR = 5 dB  & 42.18 & \textbf{12.06} & 56.66 & \textbf{6.37} & 71.20 & \textbf{13.92} \\
                \midrule
                \multirow{4}{*}{Pink Noise}
                    & SNR = 20 dB & 11.85 & \textbf{9.10}  &  6.58 & \textbf{4.38} & 46.40 & \textbf{13.91} \\
                    & SNR = 15 dB & 13.83 & \textbf{10.42} &  7.90 & \textbf{4.42} & 48.49 & \textbf{16.29} \\
                    & SNR = 10 dB & 15.00 & \textbf{11.77} &  9.91 & \textbf{4.55} & 48.92 & \textbf{17.82} \\
                    & SNR = 5 dB  & 16.24 & \textbf{12.45} & 13.82 & \textbf{4.88} & 50.09 & \textbf{18.47} \\
                \midrule
                \multirow{4}{*}{Brown Noise}
                    & SNR = 20 dB & 11.87 & \textbf{8.19}  &  6.66 & \textbf{3.93} & 46.98 & \textbf{20.28} \\
                    & SNR = 15 dB & 13.90 & \textbf{9.24}  &  7.89 & \textbf{3.94} & 48.06 & \textbf{23.59} \\
                    & SNR = 10 dB & 14.92 & \textbf{10.40} &  9.86 & \textbf{3.88} & 49.29 & \textbf{26.18} \\
                    & SNR = 5 dB  & 16.16 & \textbf{10.95} & 14.10 & \textbf{4.05} & 50.67 & \textbf{26.39} \\
                \midrule
                \multirow{2}{*}{Telephone}
                    & 300--3400 Hz & 52.49 & \textbf{12.62} & 49.53 & \textbf{16.67} & 48.10 & \textbf{1.88} \\
                    & 500--3000 Hz & 50.65 & \textbf{13.44} & 52.35 & \textbf{18.91} & 34.94 & \textbf{2.05} \\
                \bottomrule
            \end{tabular}
        \end{table}

\section{Conclusions}\label{sec:conclusions}
     The paper presents an explainable-by-design framework for audio deepfake detection based on Wiener-Hopf linear prediction.
    Unlike conventional black-box architectures, the proposed approach employs physically grounded and interpretable features processed by a lightweight CNN classifier, enabling a direct and transparent link between detection decisions and the acoustic properties of the signal.
    The proposed method achieves competitive detection performance across heterogeneous benchmarks while maintaining a significantly lower computational complexity compared to state-of-the-art methods.  
    
    Interpretability analysis using Grad-CAM reveals that the classifier focuses on low-order linear prediction coefficients, as well as on silence and transition regions of the audio signal. 
    We hypothesize that this behavior reflects the ability of the Wiener-Hopf predictor to capture reverberation characteristics and subtle statistical structures that synthetic generation systems struggle to reproduce accurately. 

    Furthermore, the robustness analysis shows that fine-tuning effectively recovers detection performance under a wide range of post-processing conditions, including additive noise, MP3 compression, and telephone filtering.
    Future work will investigate domain generalization strategies and the development of other acoustic features within the explainability-by-design framework.

\begin{acks}
    This work was carried on through partnership and collaboration with the São Paulo Research Foundation (FAPESP) Horus project, Grant \#2023/12865-8.
    This manuscript reflects only the authors’ views and opinions, FAPESP cannot be considered responsible for them. All of the authors have revised the document and confirm the findings.
\end{acks}

\bibliographystyle{ACM-Reference-Format}
\bibliography{sample-base}

\end{document}